\documentclass[aps,prb,10pt,twocolumn,showkeys,showpacs,superscriptaddress]{revtex4-1}

\usepackage{graphicx}
\usepackage{dcolumn}
\usepackage{multirow}
\usepackage[breaklinks=true,colorlinks=true,linkcolor=blue,urlcolor=blue,citecolor=blue]{hyperref}
\newcommand{\comm}[1]{}

\newcommand{\hs}{H$_3$S\  }
\newcommand{\HS}{H$_3$S}
\newcommand{\TC}{T$_c$}
\newcommand{\tc}{T$_c$\ }
\newcommand{\etal}{{\it et al.}}

\begin{document}

\title{Spectroscopy of H$_3$S: evidence of a new energy scale for superconductivity}
\author{F. Capitani}
\author{B. Langerome} 
\author{J.-B. Brubach} 
\author{P. Roy}
\email{pascale.roy@synchrotron-soleil.fr}
\affiliation{Synchrotron SOLEIL, AILES Beamline, Saint-Aubin, 91190, France}
\author{A. Drozdov} 
\author{M.I. Eremets}
\affiliation{Biogeochemistry Department, Max Planck Institute for Chemistry, PO Box 3060, 55020 Mainz, Germany}
\author{E. J. Nicol}
\affiliation{Department of Physics, University of Guelph, Guelph, N1G 2W1 ON Canada}
\author{J. P. Carbotte}
\author{T. Timusk}
\email{timusk@mcmaster.ca}
\affiliation{Department of Physics and Astronomy, McMaster University, Hamilton, ON L8S 4M1, Canada}
\affiliation{The Canadian Institute for Advanced Research, Toronto, ON M5G 1Z8 Canada}

\date{\today}

\begin{abstract}
The discovery of a superconducting phase in sulfur hydride under high pressure with a critical temperature above 200 K has provided a new impetus to the search for even higher \TC. Theory predicted and experiment confirmed that the phase involved is \hs with Im-3m crystal structure. The observation of a sharp drop in resistance to zero at \TC, its downward shift with magnetic field and a Meissner effect confirm superconductivity but the mechanism involved remains to be determined. Here, we provide a first optical spectroscopy study of this new superconductor. Experimental results for the optical reflectivity of \HS, under high pressure of 150 GPa, for several temperatures and over the range 60 to 600 meV of photon energies, are compared with theoretical calculations based on Eliashberg theory using DFT results for the electron-phonon spectral density $\alpha^2$F($\Omega$). Two significant features stand out: some remarkably strong infrared active phonons at $\approx$ 160 meV and a band with a depressed reflectance in the superconducting state in the region from 450 meV to 600 meV. In this energy range, as predicted by theory, \hs{} is found to become a better reflector with increasing temperature. This temperature evolution is traced to superconductivity originating from the electron-phonon interaction. The shape, magnitude, and energy dependence of this band at 150 K agrees with our calculations. This provides strong evidence of a conventional mechanism. However, the unusually strong optical phonon suggests a contribution of electronic degrees of freedom.  

\end{abstract}


\maketitle

\comm{this is a comment, use it to make comments that will not appear in the output}
Last year, a new superconductor with a transition temperature \tc of 203 K was discovered by Drozdov \etal \cite{drozdov2015conventional}. Hydrogen sulfide, confined under the high pressure of 155 GPa in a diamond anvil cell (DAC), becomes superconducting, showing both zero electrical resistivity below \tc  and a Meissner effect. Furthermore, the superconducting phase has been found to be \hs by x-ray diffraction \cite{einaga2016crystal}. Calculations based on density functional theory (DFT) suggest that superconductivity in \hs is caused by the electron-phonon interaction, enhanced by a combination of the light mass of hydrogen and very strong coupling to high energy modes\cite{duan2014pressure,errea2015high,bernstein2015superconducts,papaconstantopoulos2015cubic,flores2016high}. What is lacking is an experimental verification of this mechanism. A step in that direction would be the identification of the spectrum of bosons that couple to the charge carriers to form the glue that leads to superconducting pairing. 
 
The mechanism whereby conventional metals become superconductors is well established and involves the electron-phonon interaction\cite{mcmillan1965lead,carbotte1990properties}. The current-voltage characteristics of a tunnelling junction\cite{mcmillan1965lead} and optical spectroscopy\cite{joyce1970phonon,farnworth1976phonon,hwang2014deriving,carbotte2011bosons} have yielded detailed information on the electron-phonon spectral density $\alpha^2$F($\Omega$) as a function of phonon energy $\hbar\Omega$. These phonon spectra were further verified by neutron scattering\cite{stedman1967phonon}. 

It is an experimental challenge to extend these methods to the recently discovered hydrogen sulfide under pressure of 150 GPa for several reasons. The sample size $\approx$ 50 $\mu$m clearly excludes neutron studies, while the DAC environment does not allow photoemission measurements. Optical studies through infrared transmission are difficult as they require thin films of the order of tens of nanometers, or less, which are difficult to make in current state-of-the-art of DACs\cite{drozdov2015conventional}. 

Reflectance measurements seem to be the best alternative in such high pressure environment\cite{perucchi2009optical}. Nevertheless, several problems remain to be resolved, which result from the minute size of the sample, and the presence of diamond in the optical path. To overcome these difficulties, we adopted the method of measuring reflectivity at various temperatures and then evaluated the temperature ratio between the superconducting state to the normal state. These temperature ratios were then compared to theoretical ratios from DFT and Eliashberg theory. Alternatively, for normal reflectance measurements, we used the reflectivity from either surface of the diamond (or the NaCl gasket) as a reference. 

Through these methods, we provided the evidence for the superconducting transition in \hs{} under pressure of 150 GPa and demonstrated the high energy scale involved. 

\begin{figure}[tb]
\centering
\includegraphics[width=8.8cm]{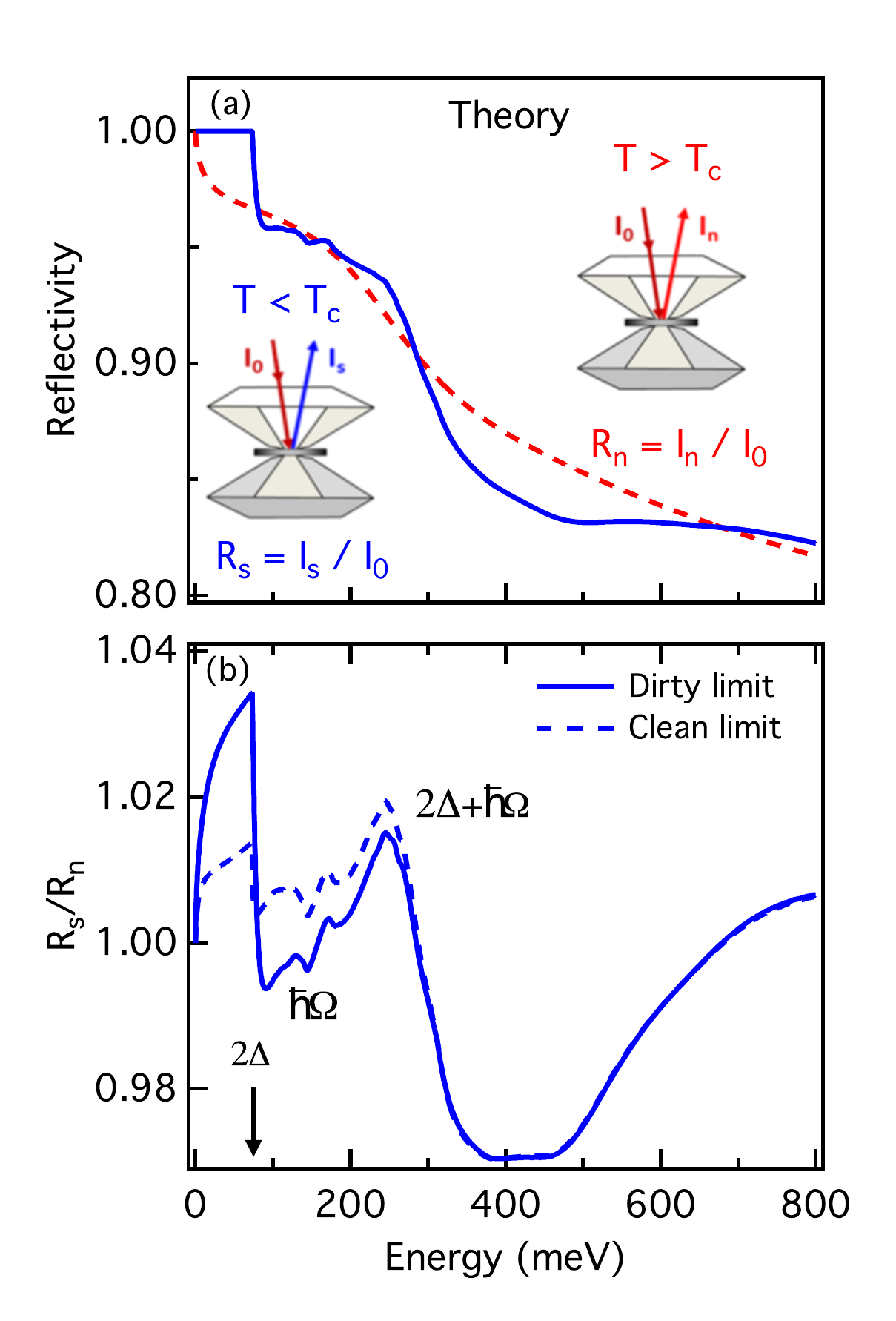}
\caption{ 
Theoretical reflectivity of \HS{}.
(a) Calculated absolute reflectivity in the normal metallic state (dashed red line) and in the superconducting one at T = 20 K (solid blue line) of \hs for $\gamma_r = 135$ meV. 
(b) Ratio of the reflectivity in the superconducting state divided by the normal state value for a sample in the dirty limit (solid blue line), \textit{i.e.} elastic scattering rate $\gamma_r = 135$ meV, and the same ratio for a sample in the clean limit (dashed blue line), $\gamma_r = 28$ meV. 
Note the three regions where the reflectivity in the superconducting state differs from the metallic: (i) The superconducting gap is predicted to give rise to a sharp step of 3.6 \% at 73 meV in the dirtier sample and a weaker step for the clean sample; (ii) in the 100-200 meV region, direct absorption by phonons can give rise to absorbing features, as indicated by $\hbar\Omega$; (iii) in the 300-500 meV region, a strong depression at energies between $2\Delta + \hbar\Omega$ and $2\Delta + 2\hbar\Omega$ is due to scattering from bosonic excitations, with a recovery at larger energy values.}
\label{Rs_Rn}
\end{figure}

Fig. \ref{Rs_Rn}(a) shows the calculated reflectivity in the normal metallic state $R_n$ of \HS, as the dashed red curve, and in the superconducting state, $R_s$ as the solid blue curve. The details of these calculations are given in the Supplementary Information (SI). As shown in Fig. \ref{Rs_Rn}(a), a superconductor with an isotropic gap is a perfect reflector of electromagnetic radiation, {\it i.e.} $R_s = 1$, up to a photon energy $\hbar\omega = 2\Delta$. Above this energy, the reflectance drops to approximately the normal metallic state value. 

Fig. \ref{Rs_Rn}(b) shows the calculated reflectance ratio, $R_s(T)/R_n$ for two values of the elastic scattering rate $\gamma_r$, one sample is near the clean limit while the other one is closer to the dirty limit (see the following section and the SI). The gap ($\Delta$) gives rise to a step at $2\Delta= 73$ meV in the calculated reflectivity\cite{nicol2015comparison}. This drop is larger if the sample is closer to the dirty limit, {\it i.e.} if the scattering rate is larger than the gap: $\gamma_r \gg 2\Delta$ \cite{kamaras1990clean}. Such effect is directly translated into the reflectivity ratio, as illustrated in Fig. \ref{Rs_Rn}(b), where a $\sim$ 3\% drop of $R_s(T)/R_n$ is expected at 73 meV in the dirty limit whereas it is reduced to $\sim$1\% in the clean limit.

In the following, we will focus our investigation to a sample in the dirty limit measured over three regions: the gap region, 65 to 100 meV, the range of direct absorption by optically active phonons, 100 to 200 meV, and the boson region, 450 to 600 meV, where the reflectance ratio is depressed by some 3.5 \% at 450 meV and rises between 450 and 800 meV as the energy increases. This depression of reflectance is due to the strong scattering from bosonic excitations with $\hbar\Omega \approx 200$ meV.\\

\begin{figure}[tb]
	\centering
	\includegraphics[width=8.8cm]{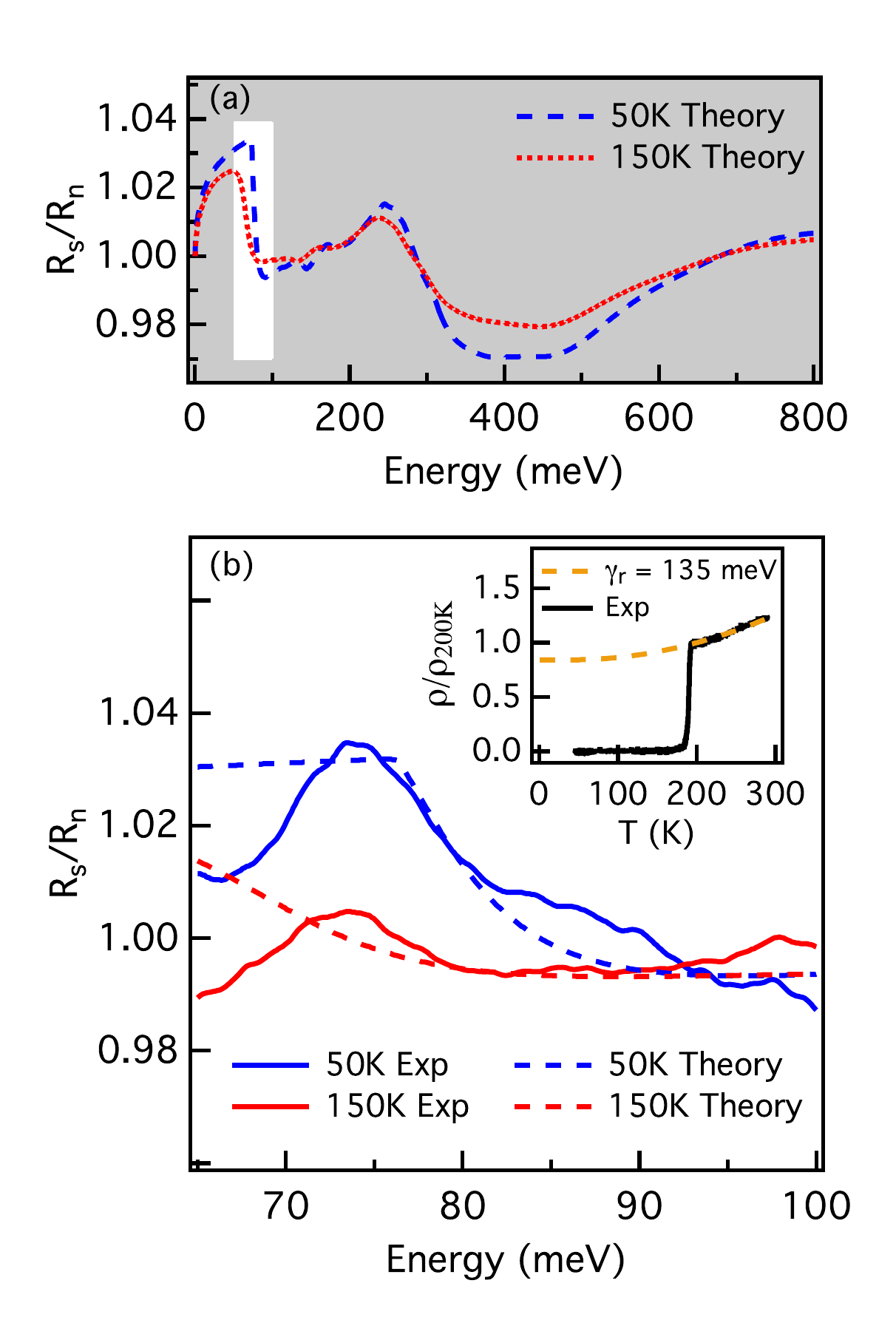}
	\caption{
		The superconducting gap of \HS. 
		(a) The curves show the theoretical reflectivity ratio in the superconducting state divided by the normal state value $R_s(T)/R_n$ for a sample in the dirty limit at 50 K (dashed blue line) and at 150 K (dotted red line). The white area highlights the energy region where the measurements presented in panel (b) were performed.
		(b) The blue and red solid curves indicate the measured $R_s(T)/R_n$ for $T=$ 50 K and 150 K, respectively. The drop of the signal below 70 meV is due to instrumental artifacts. The corresponding calculated reflectivity ratios are shown as dashed curves. The inset presents the DC resistivity (black curve) of the same \hs sample. Data have been normalized to unity at 200 K. The sudden drop of resistivity at 200 K indicates the critical superconducting temperature \TC. Data between 200 and 300 K are well reproduced by a curve obtained with a residual elastic scattering rate $\gamma_r = 135$ meV, thus indicating that this sample is in the dirty limit.}
	\label{f_gap}
\end{figure}

\noindent\textbf{The energy gap}\\
Fig. \ref{f_gap}(a) shows the theoretical $R_s(T)/R_n$ for a sample in the dirty limit ($\gamma_r = 135$ meV as determined below) at two different temperatures: $R_s(150\text{ K})/R_n$ and $R_s(50\text{ K})/R_n$. These two temperatures correspond respectively to a superconducting sample close to $T_c$ (150 K) and one in a well-established superconducting state (50 K). Here, the drop in reflectance ratio is larger when temperature lowers and the onset of this structure shifts from 50 meV for T = 150 K to 73 meV at 50 K.  

Fig. \ref{f_gap}(b) shows our measured $R_s(T)/R_n$ for a sample of \hs determined to be in the dirty limit at two temperatures compared with the corresponding theory curves. Data are limited to the energy region where the superconducting gap is expected (as highlighted in Fig.\ref{f_gap}(a)).

For this sample, the DC resistivity was used to determine \tc and the elastic scattering rate $\gamma_r$. The temperature dependence of the resistivity is shown in the inset of Fig. \ref{f_gap}(b), where an abrupt drop indicates a superconducting critical temperature \tc$\approx$ 200 K. 
Note that the measured resistivity is not quite linear in temperature above 200 K but rather increase by an additional 20 \%. This provides a first indication that the bosons involved in the inelastic scattering have high energies since a deviation from a linear $T$ law is seen only at low temperatures in ordinary metals.
The resistivity above \tc is well described by the dashed curve (shown in orange in the inset of Fig. \ref{f_gap}(b)) obtained with $\gamma_r$ = 135 meV thus confirming the dirty limit nature of this sample (see SI for further details).

As shown in Fig. \ref{f_gap}(b), we observe a clear feature at 76 meV that is present in the superconducting state only. The measurement at 50 K shows a ~3\% increase at 76 meV while the measurement at 150 K, closer to $T_c$, shows a modest increase slightly above the limit of our experimental sensitivity of about 1\%. The intensity of this feature and its temperature dependence are in good agreement with the theoretical behavior of the superconducting gap (see dashed curves).

Notice, however, that there are a few discrepancies between our observed gap signature and the predicted one: (i) - the theory predicts a step-like feature while our experimental curve is peak-like; (ii) - according to theory, as the temperature is increased, the gap is expected to close, while the gap feature gets weaker but does not shift significantly. We suggest these effects to be caused by a temperature dependent absorption by the TO-LO reststrahlen band of NaCl shifted to higher frequencies by pressure (NaCl is used as a gasket in the DAC). The narrowing of this band as the temperature is lowered could explain the strong reduction of signal at lower temperature in the 60 to 70 meV region. A run with a larger spot showed an increase in the amplitude of these features (further details can be found in the SI). Finally we note that while our theoretical model predicted a gap at 73 meV, our data fits a somewhat higher value of 76 meV. \\

\noindent\textbf{Structure from bosons}\\
The straightforward technique to extract the boson spectrum of a new superconductor is through an analysis of the absolute reflectance in the energy region $\hbar\omega \approx 2\Delta + \hbar\Omega_{log}$, where $\Delta$ is the superconducting gap and $\Omega_{log}$ is a characteristic boson energy. Such measurement is difficult to realize in a high pressure system. As we have done for the gap structure measurements, we adopt the method of measuring reflectance ratios using the normal state reflectance as a reference and comparing these ratios with calculations from DFT and Eliashberg theory.  

\begin{figure}[htb]
	\centering
	\includegraphics[width=8.8cm]{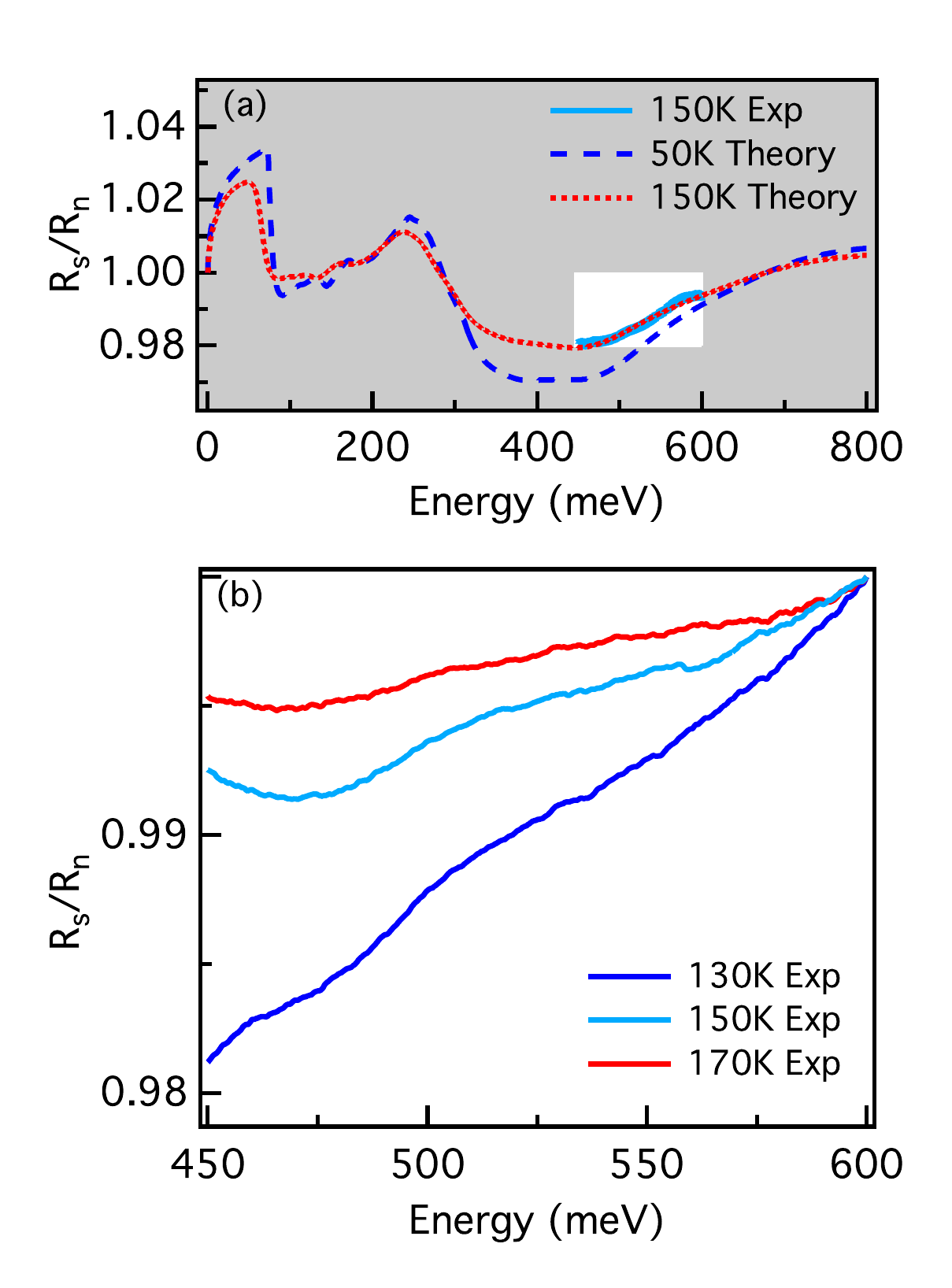}
	\caption{ 
		The bosonic signature of \HS. 
		(a) The curves show the calculated $R_s(T)/R_n$ for a sample in the dirty limit at 50 K (dashed blue line) and at 150 K (dotted red line). The light blue line corresponds to a measured $R_s(T)/R_n$ at 150 K averaged over four runs. The white area highlights the energy region where the measurements presented in the lower panel were performed.
		(b) The red, light blue and dark blue lines show $R_s(T)/R_n$ at selected temperatures for a single run. Note the decrease of the slope as the temperature approaches $T_c$. All curves have been normalized at 600 meV.}
	\label{boson1}
\end{figure}

Fig. \ref{boson1}(a) shows that the main feature in the ratio is the strong 3.5 \% and 2.5 \% drop at 50 and 150 K, respectively, just above the 250 meV range.  Unfortunately, this drop is in the region of strong phonon absorption in the diamond anvil and cannot be accessed for accurate measurements. Instead, in what follows, we focus on the the slope of $R_s(T)/R_n$ in the somewhat broader region 450 to 600 meV, as highlighted in white in Fig. \ref{boson1}(a). The theory gives a slope of $1.05\times 10^{-4}$ meV$^{-1}$ at 200 GPa and $1.5\times 10^{-4}$ meV$^{-1}$ at 157 GPa.  Our experimental value for this slope is $1.07\pm0.2\times 10^{-4}$ meV$^{-1}$ at 150 GPa, in reasonable agreement with the calculations.

The measured reflectance ratio $R_s(T)/R_n$ at selected temperatures (solid lines), and therefore the temperature dependence of this increase, are shown in Fig. \ref{boson1}(b)\comm{, together with theoretical calculations (dashed black curve)}. In fact, one can see that for the measurement at T = 170 K, the ratio is 0.995 at 450 meV and rises by only 0.005 over the range. Indeed, as the temperature increases, the slope of the feature decreases and the measured reflectance approaches unity.

This behaviour indicates that superconducting \hs becomes a better reflector in the multi-boson region as the temperature is increased, in agreement with Eliashberg calculations (see SI, Fig. 2S). This effect and the good agreement of theory and experiment for the slope of $R_s(T)/R_n$ in the region 450 to 600 meV thus demonstrate that \hs is an Eliashberg superconductor, driven by the electron-phonon interaction with strong coupling to high phonons of order of 200 meV.\\

\begin{figure}[!t]
\centering
\includegraphics[width=8.8cm]{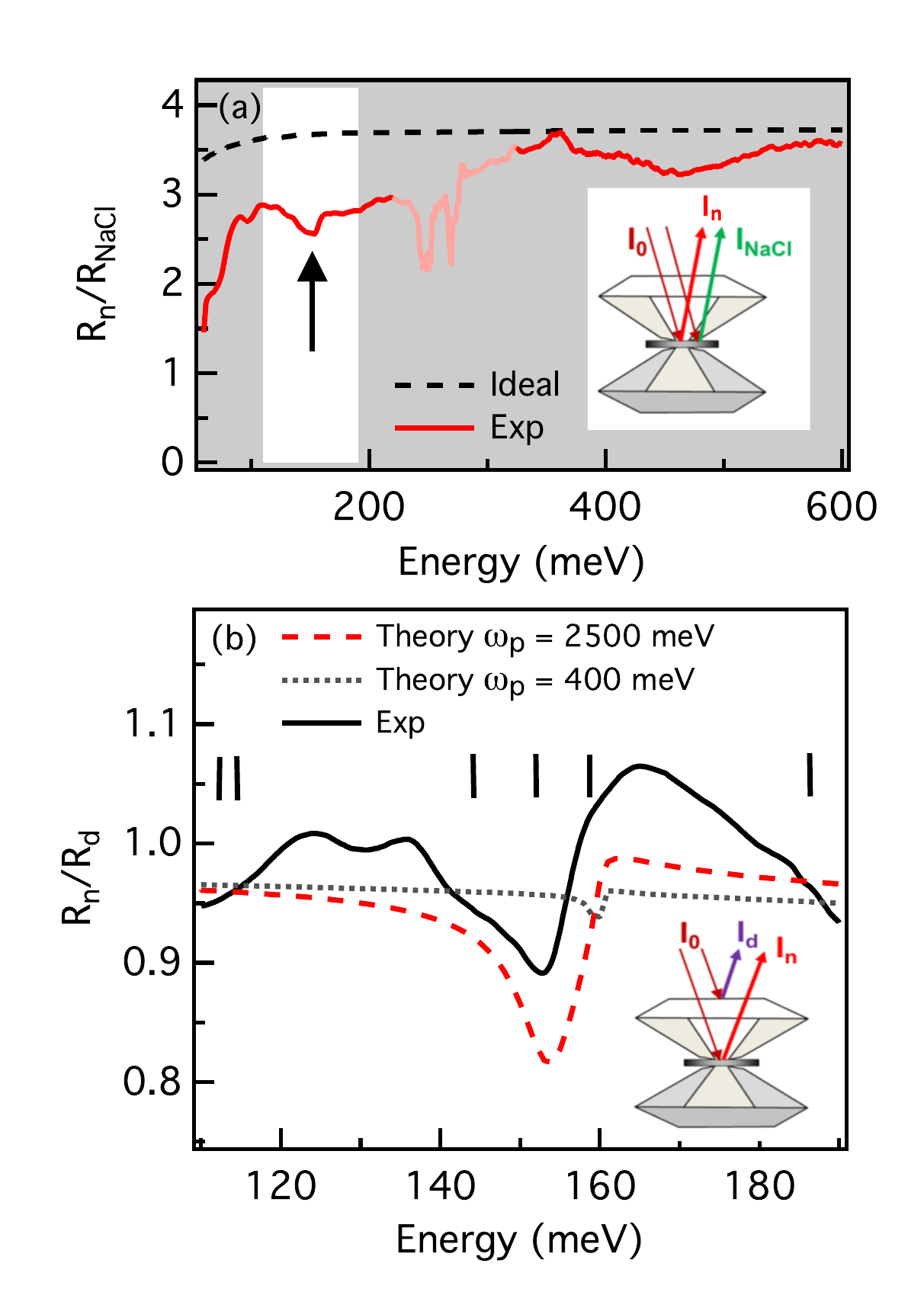}
\caption{ 
Phonon structures in the reflectance of \HS. 
(a) The measured reflectance of \HS{} in the normal state, evaluated using the NaCl gasket as the reference ($R_n/R_\mathrm{NaCl}$), is presented in a large energy range. The white area highlights an energy region where a strong phonon absorption is observed, indicated by an arrow. The red pale portion of the spectrum around 250 meV corresponds to the diamond absorption range where measurements are not possible. The drop of the reflectance ratio to 1.5 at low frequency is due to the restrahlen band of NaCl. The dashed black curve is the calculated reflectance of a perfect reflector divided by the NaCl reference at ambient pressure. A constant correction factor was applied to this curve to compensate for the smaller sample (size $\sim$ 50 $\mu$m) while for NaCl, the light is reflected from the complete spot size ($\sim$ 70 $\mu$m).
(b) The measured reflectance of \hs in the normal state at 300 K, evaluated using the front surface of diamond as reference (black curve), is shown in the phonon region. The vertical marks represent the predicted energies of \hs phonons according to Errea \etal \cite{errea2016quantum}. The calculated reflectance of \hs with an optically active phonon centered at 160 meV and a plasma energy $\omega_p = 400$ meV is shown as a grey dotted curve, while the red dashed curve is obtained with an enhanced plasma energy of 2500 meV.}
\label{NaCl}
\end{figure}

\noindent\textbf{Optically active phonons}\\
In order to estimate the absolute reflectance of \HS, we exploited the NaCl gasket surrounding the sample as a reference, as illustrated in the inset of Fig. \ref{NaCl}(a). Although its reflectance level is low, it has the advantage of having a constant reflectance for energies above its TO-LO phonon band. The TO-LO phonon energy of NaCl at ambient pressure is below our region of interest, however this band moves to higher frequencies as the pressure increases\cite{hofmeister1997ir}. It may therefore reach the investigated region but there are no optical constant data in the literature for this high pressure range.

The ratio of the signal from the sample in normal state to that from the NaCl gasket is shown in Fig. \ref{NaCl}(a) (solid red curve). For comparison, a ratio for a sample with an ideal unit reflectance for the whole energy range is shown as a dashed black curve. It is difficult to estimate the absolute value of the NaCl signal in our experiment for two reasons. First, the index of refraction of NaCl at this pressure is not known. Secondly, there is the effect of the beam spot spilling over the sample or the metal electrodes which, with their high reflectance, increase artificially the measured signal from NaCl. This enhancement of the NaCl reflected signal is expected over the whole energy range. 

\comm{The negative slope could be attributed to the broadening of the focal spot by diffraction: the spillover increases at low frequencies. Together, these two effects mask the wide range frequency dependence of the \hs signal.}

Despite these effects, several spectroscopic features stand out in Fig. \ref{NaCl}. The most striking is a strong absorption band at 150 meV. We attribute this feature to a group of optically active phonons in \HS, predicted to occur in this energy range by Errea \etal \cite{errea2016quantum}. Our models show that they are expected to cause a decrease in reflectance of \HS. No other spectroscopic signatures in the 100 to 600 meV range are observed, except a broad band from 350 to 550 meV, perhaps caused by a structure in the \hs density of states near the Fermi level or by interband transitions. Finally, one can see a rapid drop in the reflectance ratio below 100 meV which might come from the TO-LO band of NaCl, shifted to higher energies due to the high pressure. Notice that the diamond phonons absorption is cancelled out since both the sample and reference beams travel through the same thickness of diamond.

Using the diamond front surface as reference, we also derived the parameters of the phonons at $\sim$ 150 meV and compared them with theoretical predictions in terms of energy and intensity. The calculated and measured phonon absorptions are presented in Fig. \ref{NaCl}(b). Here, $I_n$ is the signal from the beam focused on the sample and $I_d$ from the beam on the front surface of the diamond (see inset). Just as with the NaCl reference, a strong asymmetric depression of reflectance with a minimum at 153 meV is apparent. This feature is temperature independent and persists up to room temperature. As for the evaluation using NaCl reference, this ratio is not an exact measure of the sample reflectance due to possible spillover, however we can estimate the relative value of the drop due to phonon structure to be 0.11 $\pm$ 0.02 (see SI for details).   The measured reflectance can be nicely reproduced (drop of 0.14) by a model conductivity that includes two components, a Drude peak with a plasma frequency of 14 eV and a damping calculated from the DFT bosonic spectral function  (see SI for details), and a Lorentz oscillator with a center energy of 160 meV, a plasma energy of 2500 meV and a damping of 1.0 meV. Fig. \ref{NaCl}(b) also presents the expected reflectance from this model along with another model with a plasma energy of only 400 meV. This lower plasma frequency would be consistent with what is expected for TO phonons in \HS. It is obvious that this latter phonon model does not fit the data. These comparisons suggest that phonons in \hs are coupled to electronic degrees of freedom, causing a remarkable increase in their strength.\\
 
\noindent\textbf{Conclusions}\\
This optical study, combined with DFT and Eliashberg theory, provide evidence for the conventional superconducting mechanism in \hs{} under pressure and demonstrate the high energy scale involved.

An unexpected result of our work is the observation of optically active phonons at 160 meV and their dramatically enhanced oscillator strength. This quasi electronic strength enhancement may be related to the anharmonic potential found by Errea \etal \cite{errea2015high}. Such ``charged phonons" where lattice vibrations acquire electronic oscillator strength have been reported in organic conductors and C$_{60}$\cite{rice1976organic,rice1992charged}. 

We observed a feature at 76 meV in good agreement with the theoretical gap structure. We also showed evidence for the predicted boson depression of reflectance in the 450 to 600 meV region in going to the superconducting state. This high energy scale is in accord with the model where the pairing glue comes from a phonon with a characteristic energy of order of 200 meV. 

Therefore, we have presented spectroscopic evidence that the superconductivity mechanism in \hs is the electron-phonon interaction with a bosonic spectral function $\alpha^2F(\Omega)$ as calculated from DFT and Eliashberg theory. \\

\begin{figure}[htb]
	\centering
	\includegraphics[width=8.8cm]{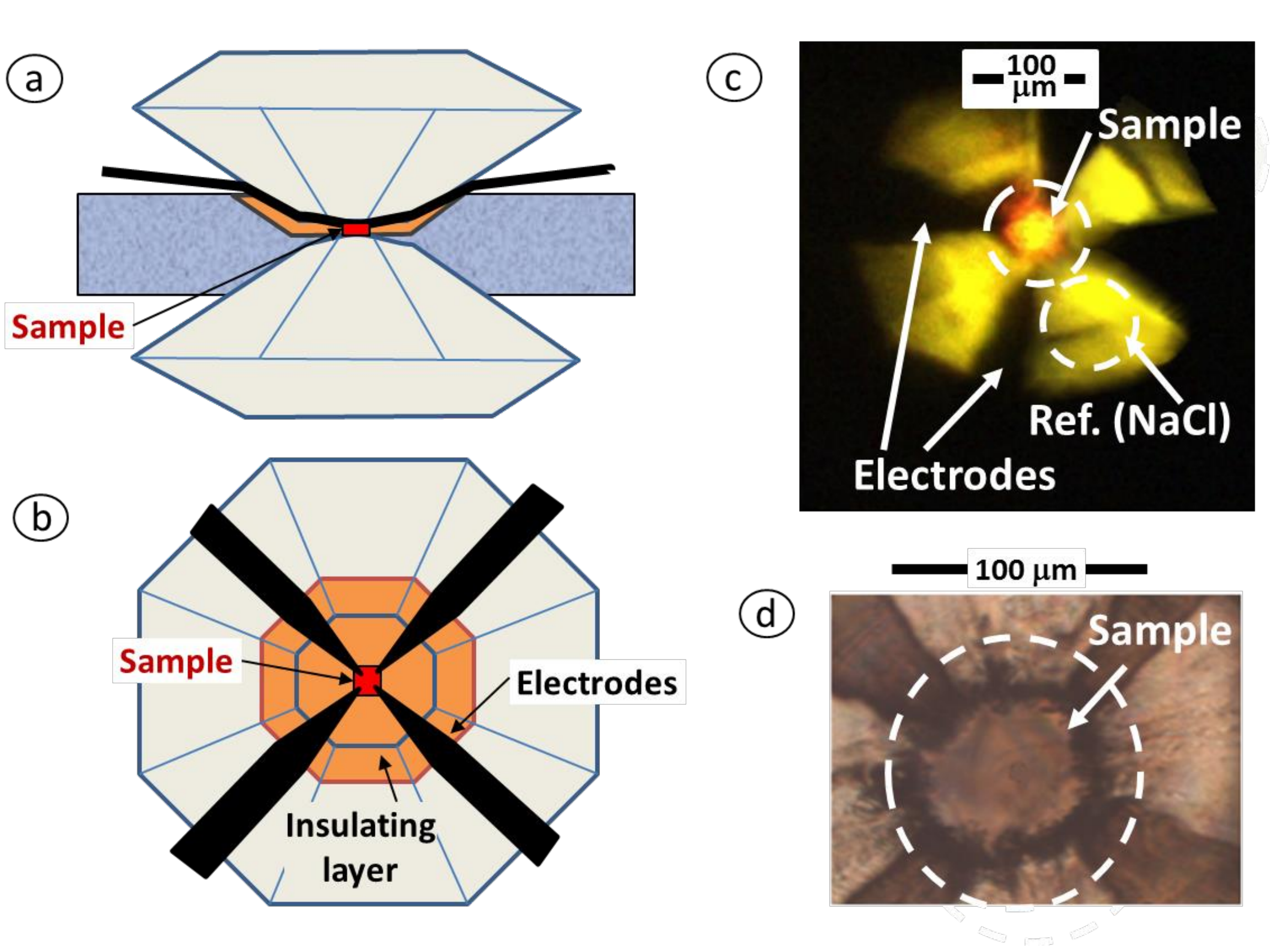}
	\caption{Diamond Anvil Cell
		(a)-(b): Schematic view of the used diamond anvil cell (DAC) showing the diamond anvils, the insulating NaCl gasket, the sample and the four electrodes for the resistivity measurements.
		(c) Camera picture of the visible synchrotron radiation focused on and reflected from the sample (orange spot in the center). The dashed circles correspond approximately to the spot size in the mid infrared, which is larger than the sample causing the reflected light to include some extra contribution from NaCl or from the electrodes. The same is true when measuring the reflectivity of the NaCl gasket (used as a reference), which includes extra contributions from the \HS{} sample. A yellow light passing through the NaCl gasket allows to visualize the cell. 
		(d) Photo of the sample inside the DAC seen through an optical microscope. The four electrodes (darker outer zones) around the sample can be also observed.}
	\label{cell}
\end{figure}

\noindent\textbf{Methods} \\
The samples were synthetized in high pressure laboratory in Mainz (Max Planck Institute for Chemistry, Mainz). H$_2$S was first confined into diamond anvil cells and annealed at 300 K to generate the \hs{} phase with a superconducting T$_c$ of $\sim$ 200 K \cite{drozdov2015conventional}. Pressure was determined using Raman spectroscopy of the diamond anvil following the method described by Eremets \etal \cite{eremets2003megabar}. This measurement was prior to all of the spectroscopic runs. Fig. \ref{cell} shows a schematic view of the cross section and a camera picture of a typical DAC used. Four electrodes allowed to measure the DC resistivity and to establish the superconducting transition temperature\cite{drozdov2015conventional}. Four samples, approximately 50 to 80 $\mu$m in diameter were investigated. NaCl was used as an insulating gasket material.

The reflectance measurements were carried out at the AILES beam line at the SOLEIL synchrotron \cite{roy2006ailes}, using a setup specially designed for measurements of samples at high pressure and low temperature \cite{voute2016new}. The DAC was in thermal contact with the cold tip of a helium flow cryostat and could be moved along the three directions by piezoelectric drivers. A camera was used to view the radiation beam spot on the sample and to maintain optimal alignment. The reproducibility of signal from one measurement to another was $\lesssim 1\%$.The IR radiation was focused on the sample and further collected with a 15x Cassegrain objective. A Bruker IFS125HR Michelson interferometer was used with various beamsplitters and detectors to cover the energy range from 50 to 800 meV. In particular, the range 60 to 600 meV was optimally measured using a custom made He cooled MCT detector \cite{faye2016improved}.

In order to measure the absolute reflectance of \HS, we focused the IR on the NaCl gasket and collected a reference spectrum from it. The front surface of the diamond anvil was also used as reference. However, when using this latter reference, the intense phonon absorption of the diamond does not cancel out. A further difficulty was the presence of a thin layer of ice on the surface of the diamond anvil. While this layer has little effect on the spectra taken with the NaCl reference, it does affect the temperature ratio since the sublimation temperature of water is about 160 K. Therefore, the ice is absent in the normal state spectra but is clearly present in the superconducting spectra. We corrected for this effect by means of a procedure outlined in the Supplementary Information.

\noindent\textbf{Acknowledgments}\\
We thank Laurent Manceron and MBaye Faye for useful discussions and for technical guiding. JPC, EJN and TT were supported by the Natural Science and Engineering Research Council of Canada (NSERC). JPC and TT received additional support from the Canadian Institute for Advanced Research (CIFAR). BL and FC received financial support from SOLEIL synchrotron. The high pressure low temperature set-up was developed through a grant from Region Centre.\\

\noindent\textbf{Author contributions}\\
This project has been initiated and supervised by T.T., M.I.E. and P.R. Samples have been synthesized and characterized by A.D. and M.I.E. Infrared measurements and data treatment were carried by B.L., F.C., J.B.B., P.R. and T.T. The calculations were performed by E.J.N. and J.P.C. All authors contributed to the writing of the paper.\\

\noindent\textbf{Additional information}\\
Supplementary information is available online.
Correspondence and requests for materials should be addressed to T.T. or P.R.

\bibliographystyle{naturemag}
\bibliography{H3S9}

\end{document}


\title{Supplementary information\\
	Spectroscopy of H$_3$S: evidence of a new energy scale for superconductivity}
\author{F. Capitani}
\author{B. Langerome} 
\author{J.-B. Brubach} 
\author{P. Roy}
\email{pascale.roy@synchrotron-soleil.fr}
\affiliation{Synchrotron SOLEIL, AILES Beamline, Saint-Aubin, 91190, France}
\author{A. Drozdov} 
\author{M.I. Eremets}
\affiliation{Biogeochemistry Department, Max Planck Institute for Chemistry, PO Box 3060, 55020 Mainz, Germany}
\author{E. J. Nicol}
\affiliation{Department of Physics, University of Guelph, Guelph, N1G 2W1 ON Canada}
\author{J. P. Carbotte}
\author{T. Timusk}
\email{timusk@mcmaster.ca}
\affiliation{Department of Physics and Astronomy, McMaster University, Hamilton, ON L8S 4M1, Canada}
\affiliation{The Canadian Institute for Advanced Research, Toronto, ON M5G 1Z8 Canada}

\date{\today}

\maketitle

\begin{figure}[htb]
	\centering
	\includegraphics[width=8.5cm]{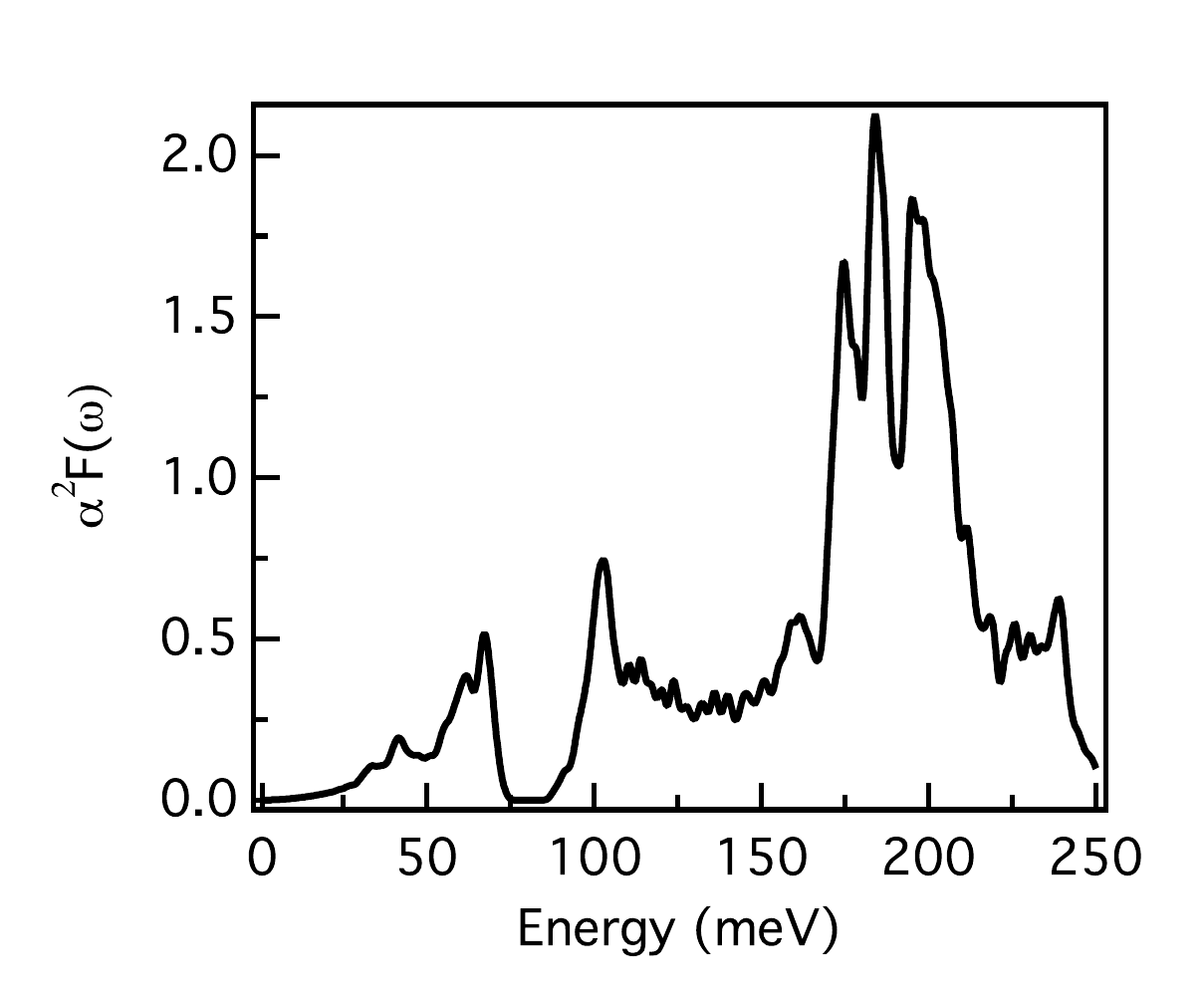}
	\caption{Electron phonon spectral density of \hs calculated from DFT by Errea {\it et al.} \cite{errea2015high}.}
	\label{a2F}
\end{figure}

\noindent\textbf{Estimation of the optical properties from DFT based $\alpha^2 F(\omega)$}\\
In this section, we describe the calculations of the expected reflectance in the superconducting and normal states of \hs shown in Fig. 1 in the main text. To accomplish this, we used the electron-phonon spectral density $\alpha^2F(\omega)$ shown in Fig. S\ref{a2F}, which has been computed by Errea {\it et al.} \cite{errea2015high} within density functional theory (DFT), providing the electronic band structure and the associated anharmonic phonons. Several other theoretical studies\cite{errea2015high,bernstein2015superconducts,papaconstantopoulos2015cubic,flores2016high} provide additional support for the general form of $\alpha^2 F(\omega)$ used here. For a value of $T_c=190$ K, Eliashberg theory gives a superconducting gap of $2\Delta=73$ meV \cite{nicol2015comparison}. While the electron-phonon mass enhancement $\lambda$ is modest and equal to 1.68 (comparable to that of Pb \cite{mori2008optical}), the spectrum extends to energies above 250 meV with significant coupling to the highest phonon energies $\Omega_{max}$. The area ($A$) under $\alpha^2 F(\omega)$ is very large in \HS, equal to 118.5 meV, more than 20 times the value in Pb for which A is 5.6 meV. We will see later that it is precisely because $A$ is so large that the reflectivity of \hs will be most favourable to measurements in the interval 450 meV to 600 meV.

What is measured most directly in our experiments is the reflectivity $R(\omega)$ (Fig. 1 in the main text), which is closely related to the optical scattering rate $1/\tau_{op}(\omega)$ and displays much the same features. The bigger the scattering rate, the more the reflectivity should deviate from unity. 
The dynamic longitudinal optical conductivity $\sigma(T,\omega)$ in the superconducting state, as a function of frequency $\omega$ and
temperature, can be calculated in Eliashberg theory\cite{errea2015high,nicol2015comparison,mori2008optical,akis1991superconducting} from the knowledge of the electron-phonon spectral density $\alpha^2 F(\omega)$ and a plasma frequency of 14 eV estimated from ref. (3).

In terms of the optical self energy $\Sigma_{op} (T,\omega)$, the generalized Drude formula is: 
\begin{equation}
\sigma(T,\omega)= i(\omega_p^2 / 4\pi)/(\omega-2\Sigma_{op} (T,\omega))
\end{equation}
 and the optical scattering rate is given by
\begin{equation}
1/\tau_{op}(T,\omega)= (\omega_p^2 /4\pi){\Re}(1/\sigma(T,\omega)).
\end{equation}

\begin{figure}[htb]
	\centering
	\includegraphics[width=8.5cm]{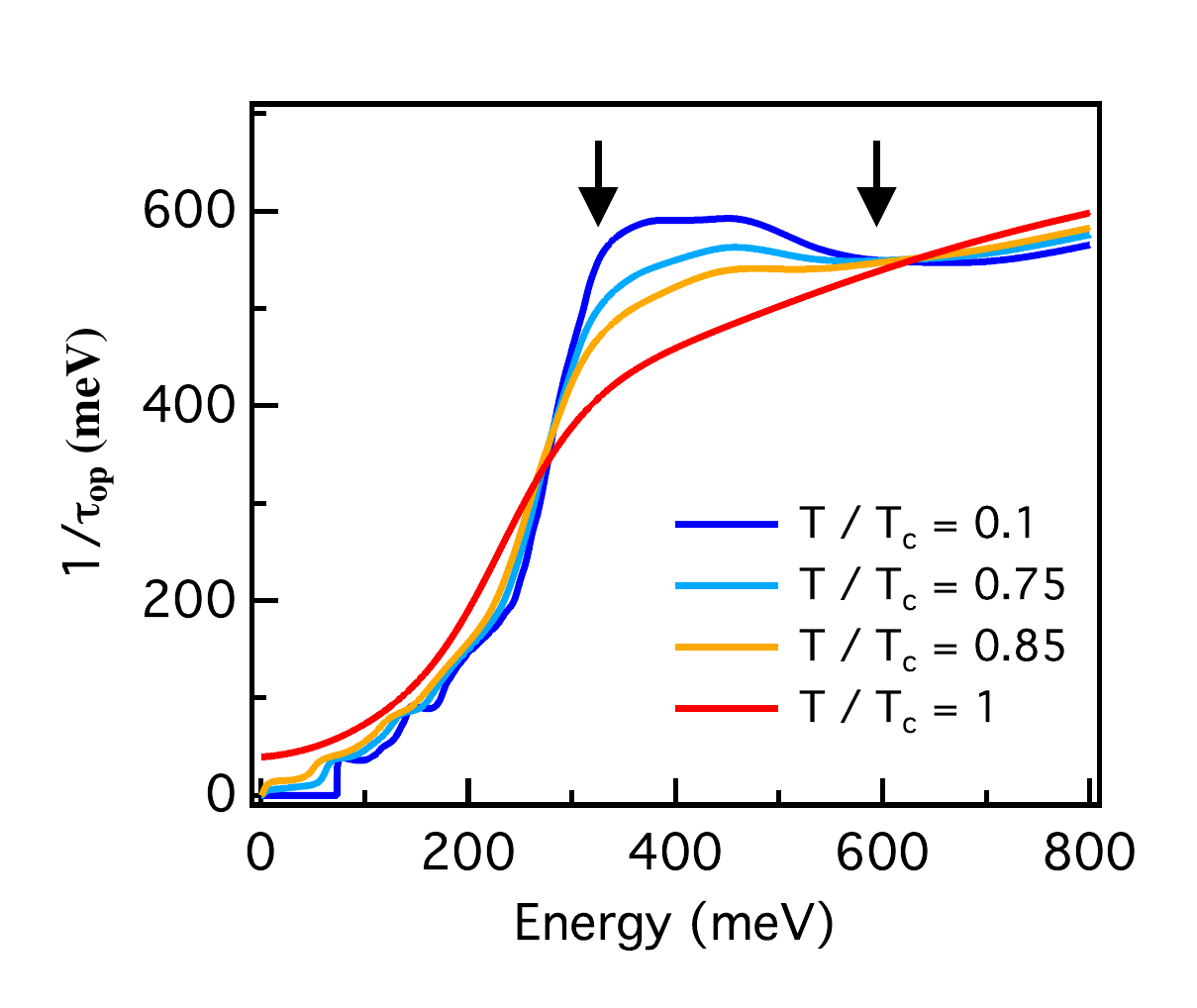}
	\caption{Optical scattering rate as function of photon energy for four temperatures. This quantity is closely related to the reflectivity, shown in Fig. 1 in the main text. The two arrows designate two characteristic energies.}
	\label{rhoT}
\end{figure}

Our Eliashberg results for the optical scattering rate of \hs in the superconducting state, at four temperatures, are presented in Fig. S\ref{rhoT}. First, note that when $T=0.1T_c$ (dark blue curve), there is no scattering below the gap at 73 meV, where a sharp vertical rise is observed. Above the gap value, an increase associated with a phonon assisted absorption is present. This evolution continues up to the first energy region, indicated by the first arrow in the figure and identified as $2\Delta+\hbar\Omega_{max}$. 

Beyond this energy scale, the scattering rate shows a flat and broad region which extends roughly up to the second energy region, marked with the second arrow and identified as $2\Delta+2\hbar\Omega_{max}$. Due to the maximal value of the optical scattering, it is clear that this energy region is the most favourable to establish the mechanism involved in the superconductivity in \hs{}. For comparison, this energy, \textit{i.e.} $\sim$ 600 meV, is higher by more than an order of magnitude than what is observed in the case of Pb. Larger scattering means larger deviations of the reflectivity from unity, which has made the present accurate reflectivity measurements possible. 
By the same argument, the low energy region will have a reflectivity which differs from unity by a small amount, thus being harder to pick up in our experiment. By contrast, in the region above the diamond absorption band, the electron-boson features have a several percent intensity. Fortunately, this energy region is the most important for uncovering the superconducting mechanism in \hs{}, should it be the electron-phonon interaction.
 
Our calculations also point out an anomalous behaviour of the scattering rate as a function of temperature within the energy region between the two arrows. Indeed, the scattering rate decreases with increasing temperature, thus making the material more reflective at higher temperatures. This is just the opposite from what is expected for an ordinary metal. The effect has its origin in the fact that in a superconductor, the quasiparticle density of states acquires energy dependence in this high energy range. When the temperature is increased, this high energy piece of the density of states decreases and consequently, the optical scattering rate drops accordingly. \\

\noindent\textbf{Reference spectrum}\\
In our attempt to measure the absolute reflectance, we used the front surface of the diamond anvil as a reference. We have to address the problem of the overlap of the focused infrared spot on the sample, which is larger than the actual sample area, phenomenon that we thereafter call spillover. Let us define $\alpha$ as the fraction of the nominal sample beam that misses the sample surface and falls on the NaCl gasket. Then, the ratio of the measured signals $r_d=I_{s}/I_{d}$, where $I_{s}$ is the signal from the beam focused on the sample and $I_{d}$ on the diamond reference, will be given by:
\begin{equation}
r_d=\frac{(R_s(1-\alpha)+\alpha R_c)(1-R_d)^2} {R_d}
\end{equation}
where $R_s$ is the reflectance of the sample, $R_c$ the NaCl gasket and $R_d$ the diamond.
In our energy range, we measure $r_d=1.0$, $R_s=0.96$ (from theory), $R_c=0.038$ (see below) and $R_d=0.172$ (from literature). We can now solve for $\alpha$ and find that $\alpha\sim0.78$. This value is reasonable since for the particular run where we measured the diamond reference with the NaCl gasket, the beam size was 100 $\mu$m and the sample was 50 $\mu$m, giving $\alpha=0.75$. In other words 75 \% of the beam misses the sample and falls on NaCl. To relate the amplitude of the features in the measured spectrum $r_d$ to the sample spectrum $R_s$, we evaluate the derivative of Eq. 1 $dr_d/dR_s$ which is given by:
\begin{equation}
\frac{dr_d}{dR_s} = \frac {(1-\alpha)(1-R_d)^2} {R_d}= 0.88
\end{equation}
with the given set of parameters. This analysis justifies in taking the measured value of $dr_d=0.11$ of the phonon feature shown in Fig. 4(b) in the main text to be close to the actual value $dR_s=0.13$. In this analysis, we neglect the absorption by the diamond which, our data show, is negligible in the 100 to 200 meV region.

In this calculation, we needed to estimate $R_c$ reflectance of NaCl interface with the diamond anvil. At ambient pressure this reflectance is 0.045 but at 150 GPa the index of refraction of NaCl and diamond has not been measured. We can extrapolate from lower pressure experiments on the dependence of index of refraction on the volume\cite{johannsen1997refractive1} and the high pressure bulk modulus\cite{ting2012isothermal} to find the reflectance of the interface dropping to 0.038 at 150 GPa.\\

\begin{figure}[htb]
	\centering
	\includegraphics[width=8.5cm]{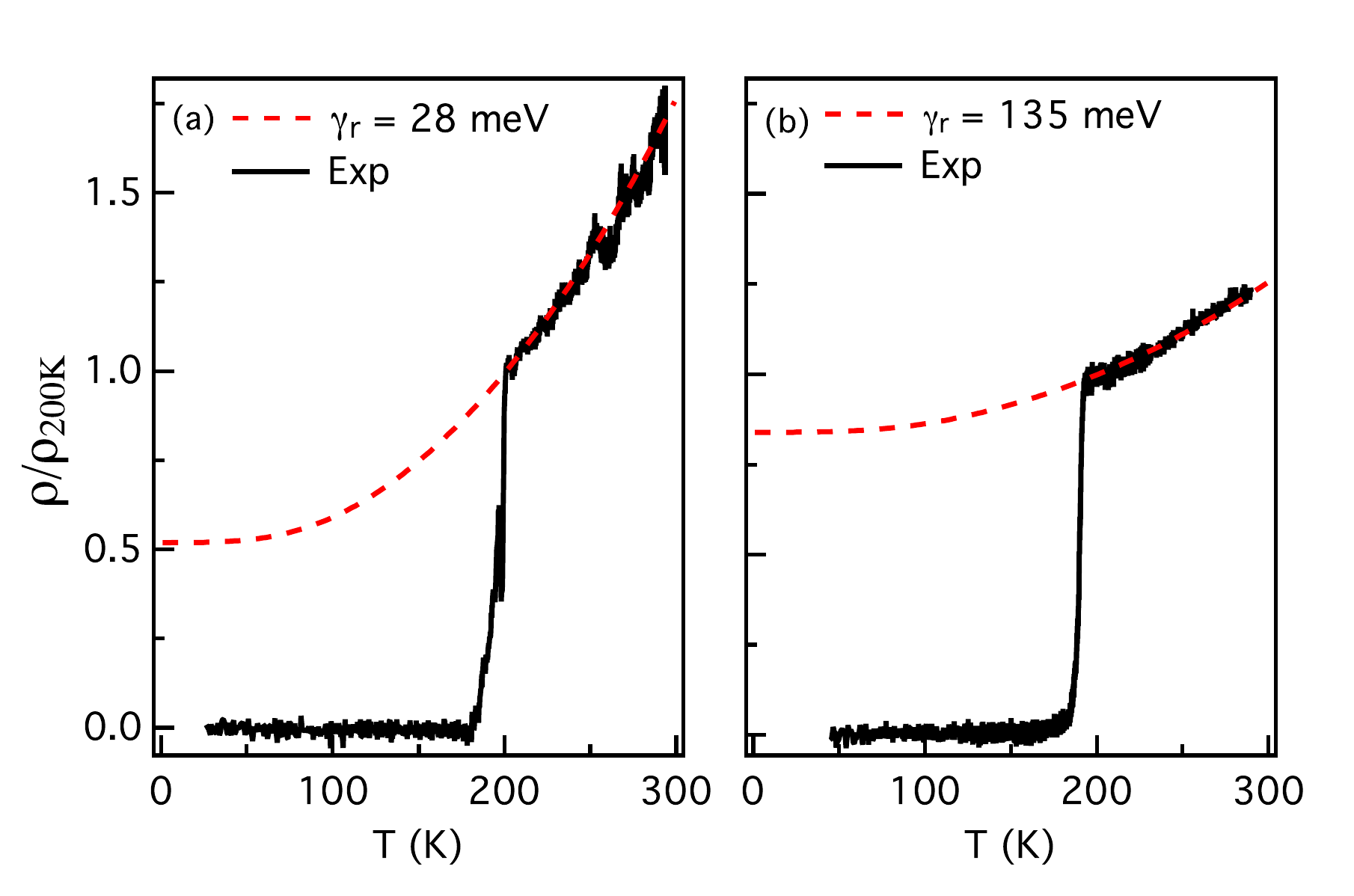}
	\caption{DC resistivity of \HS{} (solid curves) of two samples of H3S: (a) sample A and (b) sample B. The red curves show the resistivity calculated from theory, where the elastic residual scattering rate $\gamma_r$ has been chosen in order to match the experimental curve between 200 and 300 K.
		All curves have been normalized to unity at 200 K}
\label{resistivity}
\end{figure}

\noindent\textbf{Estimation of the residual scattering rate $\gamma_r$}\\
The measured DC resistivities of two samples of \HS{} as a function of temperature are shown in Fig. S\ref{resistivity} (solid lines). The resistivity data above $T_c$ can be reproduced by the dashed curves obtained from values of the elastic scattering rate $\gamma_r$ calculated with the following procedure. 

We define $\Delta R = R_2-R_1$, where $R_2$ is the resistance at 300 K and $R_1$ at 200 K, $\Delta\gamma=\gamma_2-\gamma_1$, where $\gamma_2$ is the total scattering rate at 300 K and $\gamma_1$ at 200 K, and $\gamma=\gamma_r+\gamma_i$, where $\gamma_i$ is the calculated inelastic part and $\gamma_r$ the elastic one. More specifically, we have $\gamma_1=\gamma_r+\gamma_c$ where $\gamma_c$ the calculated inelastic scattering rate at 200 K. We then have:
\begin{equation}
\frac{\Delta R}{R_1}=\frac{\Delta\gamma}{\gamma_1}=\frac{\Delta\gamma}{\gamma_r+\gamma_c}
\end{equation} 
Solving for $\gamma_r$, we find that $\gamma_r=\Delta\gamma R_1/\Delta R - \gamma_c$. By means of this equation, one finds that for typical \hs samples, $\gamma_r$ values range from 25 to 140 meV. In our case, sample A has $\gamma_r$ = 135 meV and for sample B, $\gamma_r$ = 28 meV, meaning that they are in the dirty limit and in the clean limit, respectively. A rough rule of thumb is: samples where $R_1/\Delta R \gg 1$ should be favorable for the observation of the superconducting gap. 

\bibliographystyle{naturemag}
\bibliography{H3S9}